\documentclass[prb,twocolumn,preprintnumbers,amsmath,amssymb,floats,citeautoscript,nobalancelastpage]{revtex4-1}
\usepackage{graphicx}
\usepackage{dcolumn}
\usepackage{bm}
\usepackage{times}

\begin{document}

\title{Evidence for superconducting gap nodes in the zone-centered hole bands of KFe$_2$As$_2$ from magnetic penetration-depth measurements}

\author{K.~Hashimoto$^1$}
\author{A.~Serafin$^2$}
\author{S.~Tonegawa$^1$}
\author{R.~Katsumata$^1$}
\author{R.~Okazaki$^1$}
\author{T.~Saito$^3$}
\author{H.~Fukazawa$^{3,5}$}
\author{Y.~Kohori$^{3,5}$}
\author{K.~Kihou$^{4,5}$}
\author{C.\,H.~Lee$^{4,5}$}
\author{A.~Iyo$^{4,5}$}
\author{H.~Eisaki$^{4,5}$}
\author{H.~Ikeda$^1$}
\author{Y.~Matsuda$^1$}
\author{A.~Carrington$^2$}
\author{T.~Shibauchi$^1$}

\affiliation{$^1$Department of Physics, Kyoto University, Sakyo-ku, Kyoto 606-8502, Japan\\
$^2$H. H. Wills Physics Laboratory, University of Bristol, Tyndall Avenue, Bristol, UK\\
$^3$Department of Physics, Chiba University, Chiba 263-8522, Japan\\
$^4$National Institute of Advanced Industrial Science and Technology (AIST), Tsukuba, Ibaraki 305-8568, Japan\\
$^5$JST, Transformative Research-Project on Iron Pnictides (TRIP), Chiyoda, Tokyo 102-0075, Japan }

\date{\today}



\begin{abstract}
Among the iron-based pnictide superconductors the material KFe$_2$As$_2$ is unusual in that its Fermi surface does not
consist of quasi-nested electron and hole pockets. Here we report measurements of the temperature dependent London
penetration depth of very clean crystals of this compound with residual resistivity ratio $>1200$. We show that the
superfluid density at low temperatures exhibits a strong linear-in-temperature dependence which implies that there are
line nodes in the energy gap on the large zone-centered hole sheets. The results indicate that KFe$_2$As$_2$ is an
unconventional superconductor with strong electron correlations.
\end{abstract}

\maketitle

\section{Introduction}
The discovery of high transition temperature ($T_c$) iron-based superconductors (IBS) raises fundamental questions
about origin of superconductivity \cite{Ishida09}. The microscopic pairing interactions which give rise to
superconductivity are intimately related to the structure of the superconducting energy gap which may be probed
experimentally by studying the nature of the low-energy quasiparticle excitations. In particular, the presence of nodes
in the energy gap signals an unconventional pairing mechanism, and the position of the nodes can be a strong guide as
the exact form of the pairing interaction $V_{\bm{kk^\prime}}$.

Based on analyses of spin-fluctuation mediated pairing models, several different gap structures have been proposed
\cite{Mazin08,Kuroki09,Graser09,Chubukov09,Ikeda10}. The rich variety of possible pairing states has as its origin the
unusual multiband electronic structure of the IBS. In most IBS there are disconnected quasi-two-dimensional hole and
electron Fermi-surface sheets. The former are centered on the $\Gamma$ point in the Brillouin zone and the latter at
the zone corner \cite{Ishida09}. Strong scattering between the electron and hole sheets, corresponding to a wavevector
{$\bm{q}\sim (\pi,\pi)$ leads to a nodeless gap with sign change between the hole and electron sheets (nodeless $s_\pm$
state) \cite{Mazin08,Kuroki09}. However, if in addition to this there is strong low $\bm{q}$ scattering this can
stabilize a state with nodes in the electron and/or hole bands with either $s$ or $d$-wave symmetry
\cite{Kuroki09,Graser09,Chubukov09}. $c$-axis Fermi-surface dispersion could also generate horizontal line nodes
\cite{Graser09,Craco09,Kuroki10}.

Thus far experimental studies have given evidence for two distinct types of nodal structure  in IBS \cite{Ishida09}:
one is nodeless but may have significant anisotropy \cite{Hashimoto09,Malone09,Ding08,Prozorov,Yashima09,Tanatar10} and
the other has line nodes \cite{Fletcher09,Hicks09,Yamashita09,Hashimoto10,Nakai09}. The former fully gapped state seems
consistent with the nodeless $s_\pm$ state, but it may also be explained by the conventional $s$ ($s_{++}$) state
\cite{Kontani10}. For the latter case, strong evidence of nodal lines has been obtained from the penetration depth and
thermal conductivity measurements in the compensated metals LaFePO ($T_c=6$\,K) \cite{Fletcher09,Hicks09,Yamashita09}
and BaFe$_2$(As,P)$_2$ ($T_c=30$\,K) \cite{Hashimoto10}. In both materials, recent studies
\cite{Yamashita09,Kim10,Shimojima10} suggest that the nodes are most likely located on the electron bands near the zone
corner of the Brillouin Zone, and that the hole bands centered at the $\Gamma$ point remains fully gapped.

\begin{figure}[b]
\includegraphics[width=0.9\linewidth]{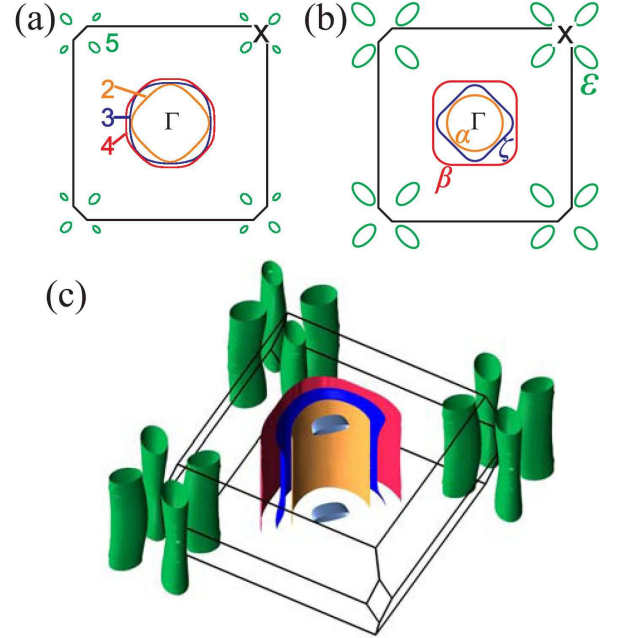}
\caption{ (Color online). The Fermi surface of KFe$_2$As$_2$ (labels in correspond to Table\:\ref{Tab1}). (a) 2D
cross-sectional representation of our band-structure calculations (note the pillow surface from band 1 does not appear
in this cut). (b) Schematic cross-section with shapes of the various pockets determined by dHvA \cite{Terashima10} and
ARPES \cite{Sato09} experiments. (c) 3D view of the DFT calculated Fermi surface with bands energies shifted to best
fit the dHvA frequencies.\protect\cite{bandenergyshifts} } \label{FS}
\end{figure}

To uncover how the superconducting gap structure is related to the microscopic pairing in IBS, studies of KFe$_2$As$_2$
may be particularly instructive. This superconductor has a relatively low $T_c \simeq 3$\,K and is the end member of
the (Ba$_{1-x}$K$_x$)Fe$_2$As$_2$ series. Unlike most other IBS where the volumes of the electron and hole sheets are
roughly equal, in KFe$_2$As$_2$ the volumes differ by one electron per unit cell \cite{Terashima10}. This causes a
substantial change in the Fermi surface topology. As shown in Fig.\:\ref{FS}, the ubiquitous $X$ centered electron
sheets are replaced by small quasi-two-dimensional hole-like tubes which do not nest at all with the $\Gamma$-centered
hole sheets. If nesting does play an important role in the superconductivity of the high-$T_c$ IBS then the nature of
the superconducting state should be quite different in KFe$_2$As$_2$. It is also intriguing that the electronic
specific heat is quite large ($\gamma=93$\,mJ/K$^2$mol) compared with other IBS \cite{Fukazawa10}, suggesting the
importance of electron correlations in KFe$_2$As$_2$. In this paper, we report measurements of the temperature
dependent penetration depth of KFe$_2$As$_2$ which shows that this material has well formed line nodes located in the
$\Gamma$-centered hole bands.

\section{Experimental methods}

The temperature dependence of the magnetic penetration depth $\lambda(T)$ was measured using a self-resonant
tunnel-diode oscillator which was mounted in a dilution refrigerator \cite{Fletcher09}. The sample (approximate
dimensions 0.2$\times$0.2$\times$0.02\,mm$^{3}$) is mounted on a sapphire rod, the other end of which is glued to a
copper block on which a RuO$_2$ thermometer is mounted. The sample and rod are placed inside a solenoid which forms
part of the resonant tank circuit which operates at $\sim$ 14\,MHz. The RF field within this solenoid is estimated to
be $<10^{-6}$\,T  so that the sample is always in the Meissner state. DC fields are screened to a similar level using a
mu-metal can.  Changes in the resonant frequency are directly proportional to changes in the magnetic penetration depth
as the temperature of the sample is varied.  The calibration factor is determined from the geometry of the sample, and
the total perturbation to the resonant frequency due to the sample, found by withdrawing the sample from the coil at
low temperature \cite{ProzorovGCA00}. The sapphire sample holder has a very small paramagnetic background signal which
varies $\sim 1/(T+\theta)$ which corresponds to a change of $\sim0.4$\,nm in $\lambda$ of our samples between 100\,mK
and 200\,mK. This is around ten times smaller than the signal from the sample and was subtracted. The $ac$ magnetic
field is applied parallel to the $c$ axis so that the shielding currents flow in the $ab$-plane. To avoid degradation
of the crystals due to reaction with moisture in the air, we cleaved the crystals on all six sides while they were
coated in a thick layer of grease. The measurements were done just after the cleavage without exposure in air. The
relatively sharp superconducting transitions found in the frequency shift of the oscillator as well as in the specific
heat measured after the penetration depth measurements [inset of Fig.\:\ref{lambda}] indicate that our procedure does
not reduce the sample quality.

\begin{figure}[b]
\includegraphics[width=0.8\linewidth]{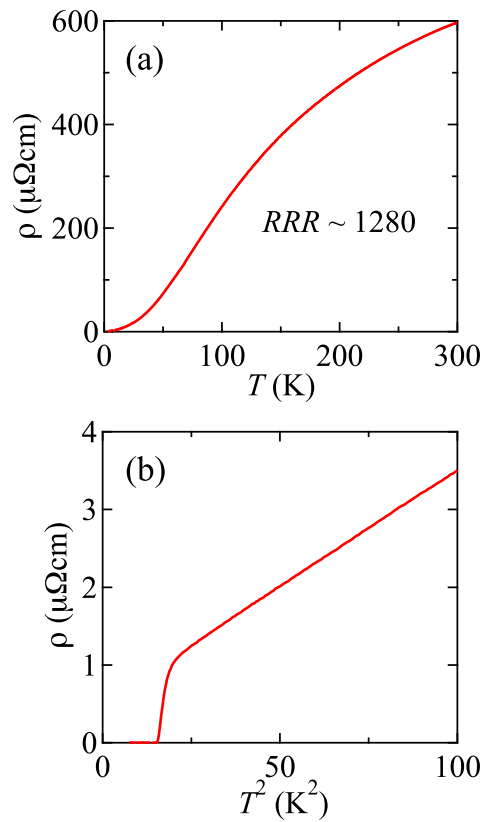}
\caption{ (Color online). (a) Temperature dependence of in-plane resistivity $\rho(T)$ in a single crystal of
KFe$_2$As$_2$. (b) The same data plotted against $T^2$ below 10\,K.} \label{rho_T}
\end{figure}

\begin{figure}[t]
\includegraphics[width=0.97\linewidth]{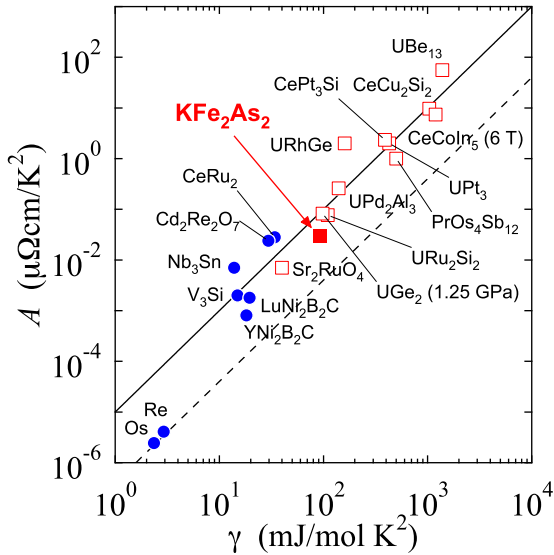}
\caption{ (Color online). Coefficient $A$ vs Sommerfeld constant $\gamma$ (Kadowaki-Woods plot \cite{KW}) for various
superconductors showing the Fermi-liquid $AT^2$ dependence of $\rho(T)$. Blue circles are for $s$-wave superconductors,
while red squares are for unconventional superconductors with nodes in the gap \cite{Kasahara06}. The lines represent
$A=a_{\rm KW}\gamma^2$, with $a_{\rm KW}=10^{-5}$ (solid line) and $4 \times 10^{-7} \mu\Omega$cm(Kmol/mJ)$^2$ (dashed
line). } \label{kw}
\end{figure}

The single crystals were grown by a self-flux method which will be described in detail elsewhere \cite{Kihou10}. The
temperature dependence of dc resistivity $\rho(T)$ was measured by the standard 4-probe method. Au contacts were
evaporated after Ar plasma cleaning of the surface, which give contact resistance less than 1\,$\Omega$. The voltage
contacts have finite widths which gives uncertainty of the absolute value of $\rho$ up to $\sim22$\%, but the
temperature dependence is not affected by this uncertainty.

In order to determine the bulk homogeneity of the superconductivity in our samples, specific heat measurements were
performed on the same single crystal sample as was used for the penetration depth measurements.  Because of the small
size of these samples (mass $\sim 4\,\mu$g), a modulated temperature method was used \cite{Carrington97}. Briefly, the
sample is glued to a 10\,$\mu$m diameter chromel-constantan thermocouple and heated with modulated light from a room
temperature LED via an optical fibre. This method has a high sensitivity but has poor absolute accuracy, so the values
are quoted in arbitrary units.

\section{Results and Discussion}
Dc resistivity $\rho(T)$ measurements show that our crystals are extremely clean with the residual resistivity ratio
$RRR=\rho(300\,{\rm K})/\rho_0$ of 1280 [Figs.\:\ref{rho_T}(a) and (b)]. The low-temperature normal-state $\rho(T)$
follows the Fermi-liquid dependence $\rho_0+AT^2$ with $A=0.030(7)\,\mu\Omega$cm/K$^2$, and the residual resistivity
$\rho_0$ is estimated by extrapolation. We note our data is not consistent with the non-Fermi-liquid $T^{1.5}$ behavior
reported by Dong {\it et al.} \cite{Dong10}, but is consistent with another previous report \cite{Terashima09}. The
obtained relatively large $A$ value follows the Kadowaki-Woods relation $A=a_{\rm KW}\gamma^2$ \cite{KW}, with $a_{\rm
KW}\approx 10^{-5}\mu\Omega$cm(Kmol/mJ)$^2$, indicating that strongly correlated electrons with large mass are
responsible for the Fermi-liquid $T^2$ dependence. As demonstrated in Fig.\:\ref{kw}, KFe$_2$As$_2$ is at the edge of
heavy-fermion superconductors with nodes in the energy gap which are widely believed to have unconventional mechanisms
of superconductivity \cite{Monthoux07}.

\begin{figure}[tb]
\includegraphics[width=0.97\linewidth]{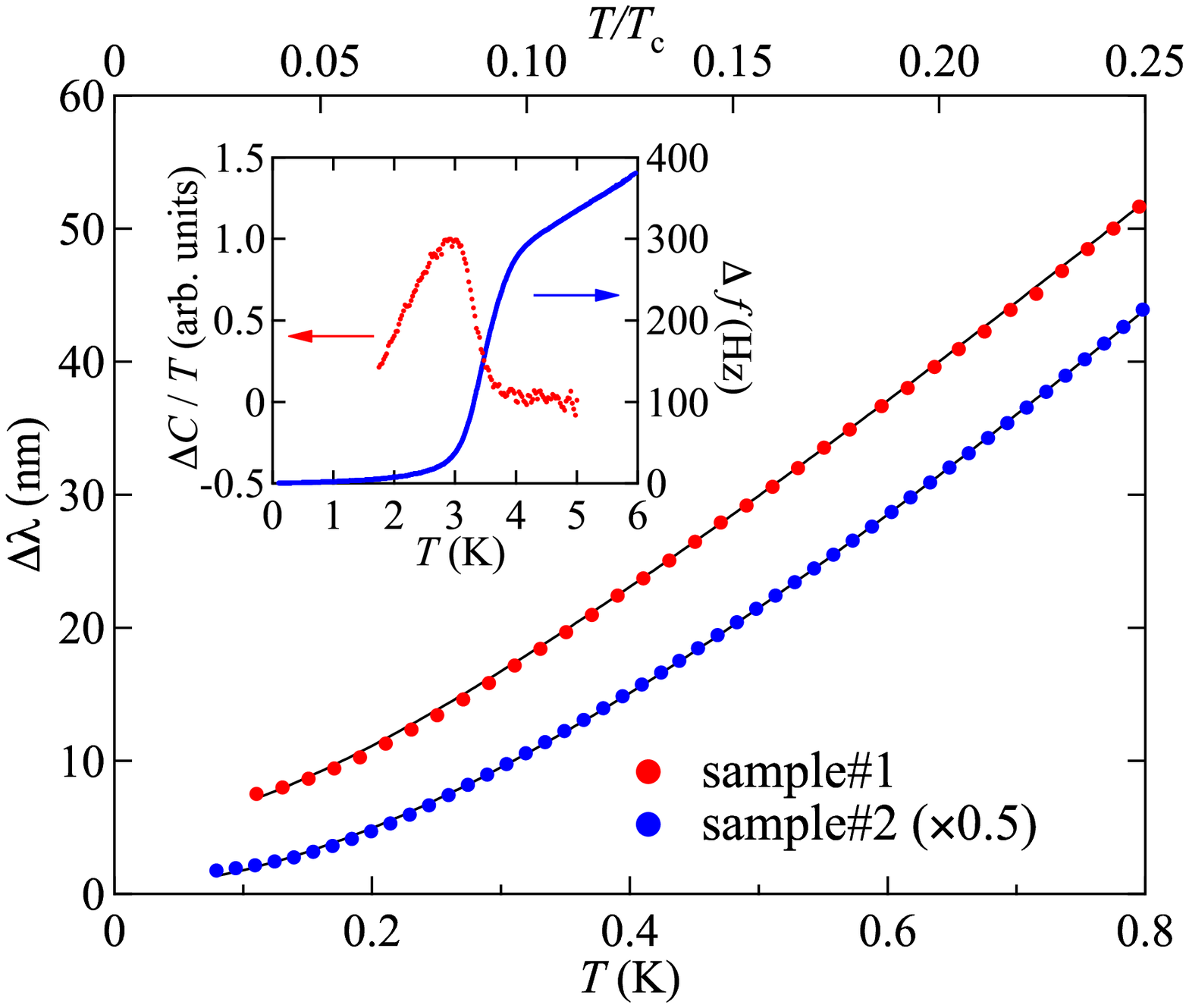}
\caption{ (Color online). Change in the penetration depth $\Delta\lambda(T)$ at low temperatures in two samples. The
data are shifted vertically for clarity. Inset shows the frequency shift $\Delta f$ of the oscillator containing sample
1, and relative change in the specific heat divided by temperature $\Delta C/T$ in the same sample.} \label{lambda}
\end{figure}

\begin{figure}[tb]
\includegraphics[width=0.97\linewidth]{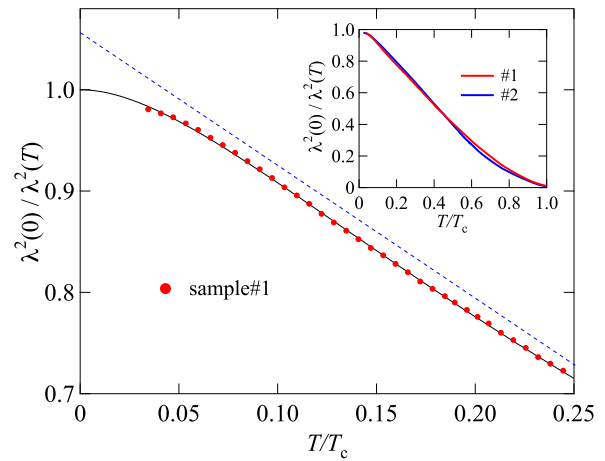}
\caption{ (Color online). Normalized superfluid density at low temperatures follows a $T$-linear dependence (dashed
line). We used $\lambda(0)=260$\,nm estimated from the Fermi-surface parameters. The solid lines are fits to the
empirical formula involving the impurity scattering in superconductors with line nodes (see text). Inset compares the
normalized superfluid density data for the two samples. To account for the factor of two difference in the slope of
$\Delta\lambda(T)$ at low temperatures, we used the doubled $\lambda(0)$ value for sample 2.} \label{superfluidlowT}
\end{figure}

\begin{table*}
\caption{\label{Tab1} Contributions of each band to the normalized superfluid density $\lambda^2(0)/\lambda^2(T)$
evaluated from DFT band structure calculations as well as the dHvA measurements \cite{Terashima10}. $m_e$ is the free
electron mass.}
\begin{ruledtabular}
\begin{tabular}{lccccccccc}
 \multicolumn{5}{c}{From DFT calculations} & & \multicolumn{4}{c}{From dHvA} \\
 \cline{1-5} \cline{7-10}
Sheet & \# holes & DOS (eV) & \multicolumn{1}{c}{$\omega_p$ (eV)} &
\multicolumn{1}{c}{$\frac{\omega_p^2}{(\omega_p^2)^{\rm total}}$ (\%)}
&  & name & \# holes & \multicolumn{1}{c}{$\frac{m^*}{m_e}$} & \multicolumn{1}{c}{$\frac{\omega_p^2}{(\omega_p^2)^{\rm total}}$ (\%)} \\
\hline
1 (pillow at $Z$) & 0.002 & 0.055 & 0.16 & 0.4 &  & -- & -- & -- & -- \\
2 (inner tube at $\Gamma$) & 0.258 & 1.11 & 1.466 & 32&  & \multicolumn{1}{c}{$\alpha$} & 0.17 & 6 & 31  \\
3 (middle tube at $\Gamma$) & 0.342 & 1.48 & 1.3959 & 29& & \multicolumn{1}{c}{$\zeta$}  & 0.26 & 13 & 23 \\
4 (outer tube at $\Gamma$) & 0.390 & 1.52 & 1.5098 & 34 & & \multicolumn{1}{c}{$\beta$} & 0.48 & 18 & 31 \\
5 (tubes near $X$) & 0.009 & 1.328 & 0.55& 5& & \multicolumn{1}{c}{$\varepsilon$} & $0.09$ & 7 & 15 \\
total & 1.00 & 5.494 & 2.577& 100 & & & 1.0 &  & 100 \\
\end{tabular}
\end{ruledtabular}
\end{table*}

Figure\:\ref{lambda} shows the low-temperature variation of the change in the penetration depth
$\Delta\lambda(T)=\lambda(T)-\lambda(0)$ in two samples. In both samples a strong $T$-linear dependence is observed
over a wide temperature range of $0.1\lesssim T/T_c \lesssim 0.25$. Such a strong $T$-linear dependence is distinctly
different from the exponential dependence expected in the fully-gapped superconducting state and is instead consistent
with gap with well-developed line nodes. A linear temperature dependence can only be explained by the presence of line
nodes \cite{Vorontsov09}. This is in contrast to power law behaviors with exponents close to 2 which could be
consistent either with nodal behavior in the dirty limit or with strong impurity scattering in e.g. the intrinsically
fully gapped $s_\pm$ state \cite{Vorontsov09}. Evidence for line nodes in this compound has also been found in NMR,
specific heat \cite{Fukazawa09} and thermal conductivity \cite{Dong10} measurements.

Deviations from the $T$-linear behavior of $\lambda(T)$ are observed at the lowest temperatures which are likely due to
a finite zero-energy density of states created by a small amount of impurity scattering. When impurity scattering is
present in superconductors with line nodes, the low-temperature $\Delta\lambda(T)$ changes from $T$ to $T^2$, which can
be described by the empirical formula $\Delta\lambda(T)\propto T^2/(T+T^*)$ \cite{Hirschfeld93}. A fit to this formula
[solid lines in Fig.\:\ref{lambda}] gives $T^*\approx0.3$\,K for sample 1 and $T^*\approx0.5$\,K for sample 2, which
indicates relatively small levels of disorder in these crystals.

Although the temperature dependence of $\lambda$ is consistent between samples the absolute values of $d\lambda/dT$
differ by a factor two.  Although our calibration procedure linking the measured frequency shift to the change in
$\lambda$ has proved to be highly accurate ($\sim$ 5-10\%) for polycrystalline elemental test samples and high-$T_c$
cuprate superconductors such as YBa$_2$Cu$_3$O$_{7-\delta}$ 
(Refs.\:\onlinecite{ProzorovGCA00,carrington01}), in some
cases where there is large surface roughness of the cut edges $\Delta\lambda$ may be overestimated. We obtain almost
identical temperature dependence of the normalized superfluid density $\lambda^2(0)/\lambda^2(T)$ if we use the doubled
$\lambda(0)$ value for sample 2 [inset of Fig.\:\ref{superfluidlowT}]. This indicates that only the calibration factor
has a factor of two differences between the two samples and the whole temperature dependence is quite reproducible. A
similar effect was found in LaFePO \cite{Fletcher09}. The lower value found for sample 1 is likely to be
more representative of the intrinsic value although in LaFePO the values of $\Delta\lambda$ found by scanning SQUID
spectrometry \cite{Hicks09} were around a factor two smaller than our lowest estimate \cite{Fletcher09}.  This has
implications for our calculations of the superfluid density as will be discussed below.

To evaluate the normalized superfluid density $n_s(T)/n_s(0)=\lambda^2(0)/\lambda^2(T)$, we need the value of
$\lambda(0)$, which we cannot measure directly in our experiment. Recent small-angle neutron scattering (SANS)
measurements \cite{Kawano10} estimate $\lambda(T=55$\,mK$)\approx200$\,nm and $\mu$SR measurements \cite{Ohishi} give
$\lambda(T_c/2)\approx 280$\,nm which also suggests $\lambda(0)\sim 200$\,nm. These values are close to the
$\lambda(0)$ values calculated from the Fermi surface parameters [see Table\:\ref{Tab1}]. Regardless of the choice of
$\lambda(0)$ value within the uncertainties, the obtained temperature dependence of $n_s(T)$ shows a $T$-linear
behavior over a even wider range of $T$ than $\lambda(T)$ itself [Figs.\:\ref{superfluidlowT} and \ref{ns}]. This is
expected because the $1-\alpha (T/T_c)$ dependence of $n_s(T)/n_s(0)$ gives $\Delta\lambda(T)/\lambda(0)= \frac{1}{2}
\alpha (T/T_c) + \frac{3}{8}\alpha^2 (T/T_c)^2+\cdots$ which leads to slightly concave (superlinear) temperature
dependence for the penetration depth as we observed.

In order to proceed with a more quantitative analysis of our results we have estimated the contribution of each of the
Fermi surface sheets to the total superfluid density.  As a first approach we have calculated the band structure of
KFe$_2$As$_2$ using density functional theory (DFT) using the \textsc{wien2k} package \cite{WIEN2k} and the
experimental lattice constants and internal positions \cite{lattice}. $4\times 10^5$ $\bm{k}$-points (in the full
Brillouin zone) were used for the calculations of plasma frequencies $\omega_p$, Fermi surface volumes and sheet
specific density of states (DOS) which are reported in Table\:\ref{Tab1}.  The Fermi surface topology and band masses
are very similar to those reported previously \cite{Terashima09} [Fig.\ \ref{FS}]. The calculated total DOS and
$\omega_p$ correspond to the Sommerfeld constant $\gamma=13.0$\,mJ/K$^2$mol and $\lambda(0)=76.6$\,nm, respectively.
The experimentally observed $\gamma=93$\,mJ/K$^2$mol (Ref.\:\onlinecite{Fukazawa10}) implies a renormalization of 7.2
in the total density of states at the Fermi level. Assuming that the superfluid density is renormalized by the same
factor leads to $\lambda(0)$ being increased to 205\,nm.   We note that in Galilean-invariant (translation-invariant)
systems like liquid $^3$He  the Fermi liquid corrections to $\lambda$ cancel in the zero temperature limit
\cite{Legget}.  However, in crystalline solids this cancellation is thought not to occur \cite{Varma} and indeed in
heavy-Fermion systems a reduction in superfluid density consistent with the thermodynamic mass enhancement is observed
\cite{Gross}.

Although the general features of this Fermi surface calculation are confirmed by angle-resolved photoemission
spectroscopy (ARPES) \cite{Sato09} and de Haas-van Alphen (dHvA) measurements \cite{Terashima10}, the exact size of the
various sheets and their warping are not. Also, dHvA measurements show that the mass renormalization effects vary
between the different sheets. So as a second approach we estimate the contribution of the various sheets to the
superfluid density directly from the dHvA measurements, assuming each sheet is a simple two dimensional cylinder. As
the largest $\beta$ sheet [see Fig.\:\ref{FS}] was not observed by dHvA, we estimate its volume from the total hole
number constraint and its mass by using the measured specific heat. Note that for sheets with more than one extremal
dHvA orbit we have taken the average.  Also we have ignored any possible contribution from the small pillow sheet (band
1).  The contribution of each sheet to the superfluid density were then calculated using
$\lambda^{-2}=\mu_0e^2n_H/m^*=\omega_p^2/c^2$, where $n_H$ and $m^*$ are the hole density and effective mass for each
sheet.  From this we obtain a total superfluid density which corresponds to $\lambda(0)=260$\,nm. The fact that this
procedure and the direct calculation from the band structure (including renormalization) give values of $\lambda(0)$
which compare favorable to the direct measurements by $\mu$SR and SANS gives us confidence in the accuracy of the
result. A key result from this analysis is that the contribution of the $\varepsilon$ band near the $X$ point is small.
It contributes only up to $\sim 15$\% of the total superfluid density.

\begin{figure}[t]
\includegraphics[width=0.9\linewidth]{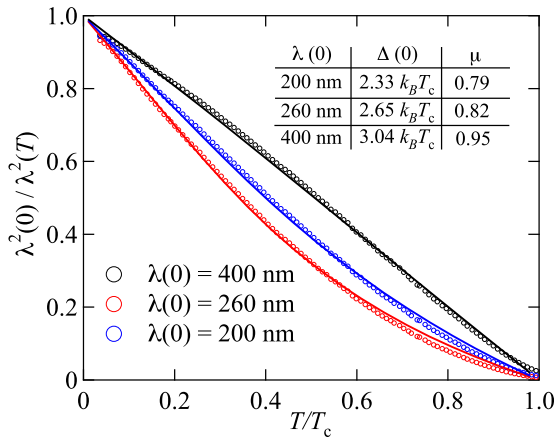}
\caption{ (Color online). Normalized superfluid density obtained by using representative values of $\lambda(0)$. The
lines are the fits to the nodal-gap model with parameters indicated in the figure.} \label{ns}
\end{figure}

The temperature dependence of the normalized superfluid density $\rho_s=\lambda^2(0)/\lambda^2(T)$ is plotted in
Fig.\:\ref{ns}. Here we have used values of $\lambda(0)$ which encompass the above estimates as well as a factor of 2
larger value in case our values of $\Delta\lambda(T)$ are overestimated because of remaining surface roughness in
sample 1. These plots show that the superfluid density at $T=T_c/3$ does not exceed $\sim75$\% of the zero-temperature
value which indicates that the nodes are located on at least one of the $\Gamma$-centered bands which have large
contributions to the superfluid density.

In a nodal superconductor a rapid decrease of superfluid density with increasing temperature is indicative of a small
value of the gap slope near the nodes $\mu\Delta_0=1/d\Delta/d\phi|_{\rm node}$  (in a simple $d$-wave model\cite{Xu95}
for $T\ll T_c$: $\rho_s(T)\simeq 1-4\ln 2/\mu\Delta_0)$.  The presence of multiple Fermi surface sheets complicates the
analysis in the present case, but it is reasonable to approximately model the gaps and Fermi velocity values on the
different sheets by globally averaged values of $\mu$ and $\Delta$.  We then approximate the variation of the gap with
in-plane Fermi surface angle $\phi$ by $\Delta(\phi)=\min(\mu \Delta_0 \phi, \Delta_0)$ \cite{Xu95}. We note that for
$\mu=2$ this produces a very similar form of $\rho_s(T/T_c)$ to the more usual $d$-wave form
$\Delta(\phi)=\Delta_0\cos(2\phi)$.  An alternative way of modelling the gap would be to add higher harmonics to this
lowest order $d$-wave form.  However, this would introduce more fitting parameters if more than one extra harmonic was
required.  In the weak-coupling limit, $\mu$ is the only free parameter in this model as the temperature dependent gap
can be calculated self-consistently \cite{Xu95}. Here however, we allow for possible strong-coupling corrections to
$\Delta_0$ and leave this as a free parameter, fixing the temperature dependence of $\Delta$ to its weak-coupling
$d$-wave form.

As shown in Fig.\:\ref{ns}, we find this model fits our data well for all assumed values of $\lambda(0)$. There is
strong covariance between the two parameters $\mu$ and $\Delta_0$ because the low $T$ slope is determined only by the
product $\mu \Delta_0$ whereas the variable ratio $\mu/\Delta_0$ influences the higher $T$ behavior only weakly.   We
find the data can also be reasonably well fitted by the inclusion of a second isotropic gap accounting for $\sim$30\%
of the total, but the large reduction of $\rho_s$ at low temperatures cannot be reproduced if we assume that only the
$\varepsilon$ tubes have line nodes as they only contribute $\sim 15$\% to the total superfluid density. This reaffirms
our conclusion that nodes must be present on the $\Gamma$ centered hole bands, although there is a possibility that
some of the sheets could be fully gapped.

As demonstrated in Fig.\:\ref{kw}, in superconductors with strong electron correlations where the $A$ and $\gamma$
values are strongly enhanced, Cooper pairs with finite angular momentum ($p$, $d$-wave, etc.) are favorable as these
states with small probabilities in small pair distances reduce the Coulomb repulsion \cite{Monthoux07}. Our results
indicate that KFe$_2$As$_2$ with $\gamma$ and $A$ values comparable to some heavy-fermion superconductors has
well-developed line nodes in the zone-centered large bands. This implies that the electron correlations play an
important role in this compound and that an electronic (non-phononic) pairing mechanism is needed to overcome the
Coulomb repulsion. Since the $\Gamma$-centered sheets in KFe$_2$As$_2$ are relatively large, it is likely that the
intra-band spin-fluctuations are important to create the sign change inside these bands, which gives rise to line
nodes. Recent theoretical calculations \cite{Ikeda10b} reveal that the spin susceptibility in heavily hole doped system
has relatively weak momentum dependence, which suggests the relative importance of scattering vectors other than
{\boldmath $q$} $\sim(\pi,\pi)$.

\section{conclusions}
In summary, our results indicating line nodes in the $\Gamma$-centered bands in KFe$_2$As$_2$ are consistent with
$d$-wave or horizontal nodal state. This is different both from the nodeless states found in (Ba,K)Fe$_2$As$_2$
(Refs.\:\onlinecite{Hashimoto09,Ding08}) and from the nodal $s$ state with nodes in the electron bands
\cite{Kuroki09,Chubukov09,Graser09} which may be the case in BaFe$_2$(As,P)$_2$ \cite{Hashimoto10,Kim10,Shimojima10}.
How this difference is linked to the changes in the Fermi surface will be an important clue towards a microscopic
mechanism of superconductivity in iron-based superconductors.

\begin{acknowledgments}

We thank discussions with D. Broun, E.\,M. Forgan, S. Kasahara, H. Kawano-Furukawa, K. Kuroki, K. Ohishi, and T.
Terashima. This work is supported by KAKENHI from JSPS, Grant-in-Aid for GCOE program ``The Next Generation of Physics,
Spun from Universality and Emergence'' from MEXT, Japan, and EPSRC in the UK.

\end{acknowledgments}




\begin{thebibliography}{99}
\bibitem{bandenergyshifts} Rigid band energy shifts of  $0$, $+20$, $-120$, $+50$, $+30$\,meV for bands 1-5
respectively were applied to best reproduce the observed de Haas-van Alphen effect measurement frequencies reported in
Ref.\:\onlinecite{Terashima10}.  The lowest frequencies ($\epsilon_l$, $\epsilon_h$ in the notation of Ref.\:
\onlinecite{Terashima10}) were assigned to the corner hole pockets (band 5).  With this energy shift the band mass of
these pockets increases to $1.7\,m_e$ so the enhancement factor $m^*/m_b=3.5$ rather than $\sim 20$ for the unshifted
bands \cite{Terashima10}.  This enhancement then close to the values for the other sheets. No orbits corresponding to
those predicted for band 1 were observed experimentally, so this was left unshifted in the figure, but in reality may
not be present.
\end{thebibliography}
\end{document}